\documentclass{ws-procs961x669}
\usepackage{graphicx,lscape}

\makeatletter
\renewcommand{\ps@plain}{%
  \renewcommand{\@oddhead}{\hfil{\footnotesize%
    A contribution to the Julian Schwinger Centennial Conference, %
    7--12 February 2018, Singapore}\hfil}%
  \renewcommand{\@evenhead}{\@oddhead}%
  \renewcommand{\@oddfoot}{\hfil\thepage}%
  \renewcommand{\@evenfoot}{\thepage\hfil}%
}
\makeatother
\crop[off]

\newcommand{\unclear}{\raisebox{0.35ex}{%
    \rule[-0.35ex]{0.4pt}{1.2ex}\underline{\scriptsize\,unclear\,}%
    \rule[-0.35ex]{0.4pt}{1.2ex}}}

\begin{document}

\title{Speeches by\\ 
       V.~F.~Weisskopf,
       J.~H.~Van~Vleck,
       I.~I.~Rabi,\\
       M.~Hamermesh,
       B.~T.~Feld,
       R.~P.~Feynman,
       and D.~Saxon,\\
       given in honor of Julian Schwinger at his 60th birthday}
\author{Berthold-Georg Englert$^a$ and Kimball Milton$^b$}
\address{%
$^a$Centre for Quantum Technologies\\%
  and Department of Physics, National University of Singapore;\\%
  MajuLab, Singapore;\\%
  {cqtebg@nus.edu.sg}\\[1ex]%
$^b$Homer L. Dodge Department of Physics and Astronomy,\\%
University of Oklahoma, Norman, OK 73019, USA\\%
{kmilton@ou.edu}}

\begin{abstract}
In February 1978 Julian Schwinger's 60th birthday was celebrated with a
SchwingerFest at UCLA. 
This article consists of transcripts of historical talks given there.
\end{abstract}

\vspace*{-1.0\baselineskip}

\section*{}
The UCLA Physics Department organized a two-day celebration of Julian
Schwinger's 60th birthday in February 1978, to which many of his former
students, collaborators, and friends came.
Scientific sessions were held during the days of the Symposium, and after the
banquet there were a number of extended remarks given by some of the
distinguished participants.
The mostly scientific talks were published in a special issue of
Physica, which also appeared as a stand-alone volume.\cite{JS60} 
The purely historical talks were deposited in the AIP archives.

The present chapter aims to make the latter talks more accessible.
They are retyped from transcriptions made at the time by one of the authors
(KAM) from no longer existent audiotapes; we present them verbatim to preserve
the freshness and vitality of the presentations.
Some extracts of these talks have appeared in Schwinger's
biography.\cite{JSbio} 

The talks by Weisskopf, Rabi, and Hamermesh were given during the plenary
sessions, while Feld's, Feynman's, and Saxon's talks (the latter with some
interjection by Rabi) were delivered after dinner.
Van Vleck was unable to come, having suffered a mild heart attack en route,
but a telegraphic message was delivered by telephone to KAM, and it was
included in Weisskopf's remarks.
Feynman's lecture was previously published in the volume celebrating
Schwinger's 70th birthday;\cite{JS70} we include it here for completeness. 

There are many wonderful stories recounted here, and it is very interesting to
see how the stories are modified when told by different participants in the
history.
The story of Victor LaMer is a case in point.
Since none of these great men are now with us, preserving their memories is
important for the future development of our science, which is as much about
personalities as it is about technical developments. 

We will not make any attempt here to offer corrections.
We can remark on a few clarifications:
(i) Since evidently titles of talks were excluded both from the program (which
has since disappeared) and the transcripts, we are uncertain which talk was
entitled ``Schwinger and the Boom in Theoretical Physics,'' referred to by
Rabi.
(ii) Herman Feshbach's talk referred to by Hamermesh appeared in Ref.\
\citenum{JS60}, p.~17.
(iii) Weisskopf, Rabi, and Saxon quoted different numbers for the quantity of
Julian's Ph.D. students.
The final official count is 73, but that includes five students receiving their
degrees after the SchwingerFest, so Saxon (and therefore Schwinger) was
essentially correct.
Of course, as noted by several, this number is largely meaningless, since
Julian Schwinger was the greatest teacher many of us have ever had, and many
justly regard him as their master.

\newpage

\section*{Victor Frederick Weisskopf (and John Hasbrouck Van Vleck)} 
Thank you for your very kind words.
It is a very special pleasure for me to be the first chairman in this, in the
celebration of the sixtieth birthday---a young boy---a man whom I love, admire
and just like.
Before becoming again sentimental, I'd like to say a few, so to speak,
administrative remarks.
You know the program unfortunately had to be changed for this afternoon
because our colleague Van Vleck has had a mild heart attack.
I am told it is nothing serious; he's not in danger, but it was not possible
for him to come here and to address you, and I would like right away for us to
read a statement of Van Vleck, which he sent by telephone, which I have here.
He says:
\begin{quote}
  ``Very disappointed to miss the ceremonial session and frustrated to be
  hospitalized so near to goal line.'' 
\end{quote}
He had to be in San Diego at that time.
\begin{quote}
 ``Abigail and I send our congratulations, not only to Julian but also to
 Clarice.'' (Applause for Mrs. Schwinger, who was present, taking pictures.)
 ``Rabi can claim that Julian is a product of New York culture,''  
\end{quote}
so he says here,
\begin{quote}
  ``but we claim Clarice as a proper Bostonian.
  Columbia is to be felicitated for its audacity and liberality in giving
  Schwinger a travelling fellowship to Wisconsin in 1937 so that he could get
  a good education right after his doctorate.
  This was the golden year in theoretical physics in Madison with Schwinger,
  Wigner, and Breit all on the campus at the same time.
  I need not elaborate on his achievement while at Harvard except to mention
  that the Karplus and Schwinger paper on line breadth is a classic which has
  been a guideline for much of my subsequent research in this field.
  I congratulate Schwinger not merely on his past research, but projected into
  the future. A few years ago I commented in the Harvard alumni magazine on
  how three former members of its faculty, Kendall, Oldenburg, and Webster,
  were still publishing at the age of 83.
  With Schwinger's sustained productivity in research displayed by his
  starting at 17, he should do at least as well and a simple calculation shows
  that he will still be publishing in 2001.'' 
\end{quote}
Now, as to the administrative matters, Herman Feshbach was so kind to agree to
speak this afternoon, though this takes away one evening for preparation.
Take this into account (laughter).

Before I now give the word to the first speaker---Rabi, naturally, who else,
could be, should be, the first speaker at a Schwingerfest---I would like to
take this opportunity as the chairman to say just a very few sentimental
words. 

I knew Schwinger, perhaps, before the War, but really I got acquainted with
him personally only when I also came to Cambridge after the end of the
Second World War in 1946.
We came to Harvard and that was, of course, an extremely exciting time.
Remember, not only because we all came back from war work, but that was the
time when the new quantum electrodynamics, the renormalization theory, Lamb
shift, magnetic moment, etc., when all this was born in these fantastic
conferences that were held---organized by Oppenheimer---in Shelter Island and
different places every year in which every conference meant a big step forward
and these big steps were, to a great extent, done by Julian himself.
But what I would like to say, a few words, I feel I have to, is the wonderful
times that we had together.
I think it was roughly on the average once a month that we had lunch together
at some strange places.
Some of them still exist.
The food was not always very good, but the conversation was good, and somehow
I, for me, to get regularly in touch with him was a very great thing, and I'm
really unhappy that he has moved to Los Angeles, although flying today, or
yesterday, from Boston to Los Angeles---Boston where there is one meter of
snow---I sort of understand at least part of the reason, but still I'm very
unhappy about it. 

Let us say a little more, because there is something which is so valuable in
having Julian around in the community of physicists and that is what I would
like to call, for lack of a better word, his style.
I find, and I'm sure that some of you will agree with me, that within physics,
within science in general, contemporary science, there is not enough variation
of style.
There is a sort of fashion style and that, of course, unfortunately is
amplified by the fact that we are forced to publish on two pages or three
pages in the terrible magazine called \textit{Physical Review Letters},
instead of having room to place leisure, so to speak, to develop style, and
therefore there's either no style or a common style, and somehow Julian has
kept and developed a style different from us, fortunately.
I'm not making a value judgment, although I could, because it is not so much
that it is a better style, but that it is a different style, which is so
necessary in physics.
But all I want to say is, that we should be grateful to Julian that he has
produced and kept and developed an individual style, and even more so strive
also to develop more individual styles so that physics becomes more colorful.
And if anybody has contributed to the color of physics in this sense, it is
Julian, and we should be grateful to him for that. 

I don't want to say more because there will be other people who can express
all this much better and I therefore hasten to give the floor to Rabi whom I
don't need to introduce. 

\enlargethispage{1.0\baselineskip}
  
\section*{Isidor Isaac Rabi}
I prepared this lecture very carefully.
I arrived yesterday.
Spent all morning.
But I left it in the hotel, so you'll be spared all that.
I once read a book review by a former professor of mine at Cornell.
In reviewing this book he said the author was at his best when he quoted, so I
will not be at my best but if you come to my hotel room~\ldots 

I don't know just where to begin since Van Vleck sent this wonderful message,
but I'll get to it.
This is an extraordinary occasion and somehow 1918 must have been an
extraordinary year.
I didn't investigate it fully, but the year 1918, separated only by about
three months, we had our birthday child here, Julian, and a little younger, in
the direction I'm pointing, is another one---this was February, the other one
born in May---Feynman, and not only were they born at this same point which is
pretty good measurement, pretty good aiming, from an experimental physicist
it's something to hit so close, but they work in the same subject, in
essentially the same way.
So in a certain sense there's still over here in the audience, we'll drop some
on Dick. 

I'm surprised we have so few people---maybe they knew the size of the room
selected, but there are a few standees.
There are a few seats back there.
Because I read last night that Julian has had over 100 Ph.D's, and this is
over a fairly long period of time.
If wives had been invited, I suppose, and children, I suppose we'd have had
the need for a very large room. 

\enlargethispage{1.0\baselineskip}

Now, when I was asked to speak, somehow I didn't know quite what to say and I
couldn't think of a topic.
Then I was called up by Mr.\ Milton and finally was pressed to give a topic,
so I gave it and it was ``Physics When Schwinger was a Boy.''
Now boy in this sense is more the western usage.
I've seen cases where you had postdocs, married, wondering how to support
family, a little gray around the edges, and yet they were referred to by
professors as ``our kids.''
So it's only in this exaggerated sense that I meant Julian was a boy.

I'm going to trace a little bit of the career of our hero.
Coincidence and fortune plays an extraordinary part in our lives.
Now Julian Schwinger was born in 1918.
In a certain sense a most unfortunate time to be born because physics was
finished at that point.
You had the great movement of Einstein, of Bohr, of Sommerfeld, Einstein
again, and the classical quantum theory was pretty well set by 1918, including
when Einstein came again.
With 1917, the famous 1917 paper, and the year before, general relativity.
So you had relativity, quantum theory, highly developed.
A theory of light, there it was, 1918.
Fortunately, the war intervened around that time.
No, the war came to an end.
It came a little earlier and the war came to an end and then there was several
years of frustration.
Well, it didn't bother Julian very much because he was busy growing up in a
beautiful part of New York City, and in 1925 came the beginning of the second
breakthrough.
The second coming of quantum, namely the beginning of the establishment of
quantum mechanics.
It didn't bother him at all, because at that time he was seven.
This was all happening.
But I was also thinking, if he had been born in time, let's say, to
participate in the great events of 1905 when Einstein was establishing
quantum, or later, in 1913, when Bohr gave the theory of the atom, and so on.
But all that passed him by.
And then came 1925 and all the great men of that period---Heisenberg, Dirac,
Pauli, Schr\"odinger, de Broglie and many, you might call minor characters,
but they were not minor by any absolute standards.
By then, Julian was going to elementary school, so things went on this way.

I can't give many details of his actual boyhood except that his biography is
very brief.
You cannot figure it out.
I didn't know when he started high school, but I was told that in his family,
he was not the bright one.
The one who got all the recognition and all the prizes was his older brother
and Julian was the dumb one---like those super \unclear, but in this case it
was rather reversed. 
But he did go to what I consider the best high school in the whole United
States.
The Townsend Harris High School, which is preparatory to City College and does
it in only three years.
And I'm told that after getting through Townsend Harris, after graduating from
Townsend Harris, college is a breeze.
What Julian did, graduate from that and he went to City College and that's
where I begin to make contact with him and I imagine the period, since he was
a very bright lad, must have been around 1934.
Is that right, Julian?
There's very little in his autobiography.
I looked it up in those Nobel Prize lectures and, whereas others have a long
tale of what they did and how they did it, Julian only describes what his aims
were, how he became interested in physics at a very early age. 

Anyway, he arrived there at City College where he did very badly.
I looked at his transcript and he was flunking English.
I said to Julian, ``Here you are flunking English. You speak very well.	What
happened?''
``I have no time to do those themes.''
Because at the time he was working on what was the real problem to him at that
time and not solved until a decade or so later---quantum electrodynamics.
Anyway, he came to my attention through Lloyd Motz, who was then teaching at
City College, and talked about this young student there, who was having great
troubles there, flunking, very bad.
I said, ``Bring him up,'' and he did bring him up and there was this child he
brought in and meanwhile---no, I was talking to Motz about the paper of
Einstein on quantum development and he said, ``Somebody's waiting outside.''
``Bring him in.''
He brought this kid in---I should say boy but boy is somehow a bad word
nowadays.
``Sit down,'' and then we started talking about the paper because this was my
way of reading a paper---taking some student and explaining it, so to speak.
At one point there was a dispute and up pipes this voice and solved the whole
problem through the sufficiency theorem.
I'm not using the right word---what you need of orthogonal functions to give
you a complete set. 
I was impressed.
So then, I said, ``What about transferring to Columbia?''
He was willing to agree, so I got his record and so on, and brought them to
admissions.
``I'd like to give this young fellow a scholarship.''
He looked at him, said, ``With this record we wouldn't admit him.''
I said something very tactless---I said, ``Suppose he were a football
player.''
I suspect it was the wrong thing to say, but I was never very tactful.
Still, the problem remained.
Just then Hans Bethe happened to come to New York and Julian showed me
something on which he was working.
It was quantum electrodynamics.
I showed it to Hans and asked, ``What do you think?''
He liked it.
I said, ``Well write me a letter.''
No man has great honor in his own country, so with this letter, Julian was
admitted and I must say he was a reformed man.
The change of location of about a mile, from 120th Street to 130th Street and
he made Phi Beta Kappa.
I never asked him closely about what happened at that point, but he did make
Phi Beta Kappa and so this is the boy part. 

Of course, what was happening in physics was beginning to sink in.
It had the revolution, the second revolution, the third, whatever they call
it, starting in 1925, and that revolution was completed more or less by 1929.
Quantum mechanics was understood.
We had the Dirac equation, transformation theories, the theory of the
electron, Pauli--Heisenberg quantum electrodynamics, Fermi's version and there
were others.
Dirac had just about shot his bolt except for the magnetic pole and the whole
thing Julian missed.
We come to the next place where the difficulties just became very great.
Quantum electrodynamics---beautiful.

But what's beautiful if it doesn't converge?
How many infinities can you live with and still call it beautiful?
So, those difficulties were there and somehow or other Julian found out about
these things.
I don't know how.
I think he found out something about it in---when he entered City College, and
he first heard then I think of the existence of quantum mechanics.
Not the old quantum theory, because I'm saying that he got his education from
the Encyclopedia Britannica.
I can't give you the volumes and numbers, but that was his mathematical
education.
When he came to City College he had heard of Heisenberg.
Julian read the book, I forget what's its name.
And also Dirac was mentioned.
So by the time he came to Columbia as a sophomore or a junior, he was very
conversant with Dirac's papers, which I think is really extraordinary.
It's not something you would think would appeal right off to a reader of the
Encyclopedia Britannica.
But he was doing very well with it.
I know I was impressed.
And as I say, he was graduated from Columbia.
His undergraduate degree and then his graduate work.
He was a great help to me because he was in my course in quantum mechanics and
whenever I had to go away, I'd ask Julian, who was an undergraduate, to take
the class.
I can assure you it was a great improvement.
He's a much better teacher than I ever was.
One of the people in that same class was Bob Marshak and it was a bad time for
being in the same class with Julian, no matter how clever you were.

Anyway, he entered graduate work and after a short time---and this is where Van
Vleck's story comes in---I thought that he had about had everything at
Columbia that we could offer.
By ``we,'' as theoretical physics is concerned, is me.
So I got him this fellowship to go to Wisconsin, with the general idea that
there were Breit and Wigner, and they could carry on.
It was a disastrous idea in one respect because, before then, Julian was a
regular guy.
Present in the daytime.
So I'd ask Julian, I'd see him from time to time, ``How are you doing?''
``Oh, fine, fine.''
``Getting anything out of Breit and Wigner?''
``Oh yes, they're very good, very good.''
I asked them.
They said, ``We never see him.''
And this is my own theory, I've never checked it with Julian, that---there's
one thing about Julian you all know, I think he's an even more quiet man than
Dirac.
He is not a fighter in any way.
And I imagine his ideas and Wigner's and Breit's or their personalities didn't
agree.
I don't fault him for this, but he's such a gentle soul, he avoided the battle
by working at night.
He got this habit of working nights---it's pure theory, it has nothing to do
with the truth. 

When he came back to Columbia after one year, a changed man.
He was taking a course with Uhlenbeck who was a visiting professor there.
Some problems of statistical mechanics, a graduate students course, but then
came time for the final examination, these oral exams.
I saw George and he said, ``What should I do about Schwinger?''
And, ``What do you mean?''
``He's taking the course and he hasn't even shown up to make arrangements for
the examination.''
``Well, that's bad.''
So I got hold of Julian and I said, ``Now look Julian, that's not polite,''
and, not a man to give offense, so he went to Uhlenbeck and arranged for an
examination.
He tried to arrange it with him for ten o'clock at night.
Well, George is no weak character.
Ten o'clock in the morning, and at this sacrifice Julian appeared.
Well George told me later, what astonished him was, not only did he answer his
questions, he'd done a little research in the field.
But what astonished him most, since he hadn't appeared in class, he did these
things in his, George's notation---by osmosis which \unclear\ anyway.

We are now talking of the period just before the war where things were going
great guns in a way---now I'm talking purely of experimental physics---things
were going great guns and Oppenheimer, the Oppenheimer school was the place to
go.
So Julian did get a National Research Council Fellowship to go there.
I thought he should go to Pauli's, but he thought Oppenheimer was a more
interesting physicist and he went there.
Spoke to Oppenheimer about it.
He said yes.
It was very good for him to have some contact with Snyder, who was one of
Oppenheimer's students or something like that, and Julian went and I spoke to
Oppenheimer later and he was terribly disappointed.
He came to the point of writing a letter to the National Research Council
suggesting that Julian go somewhere else because it took a man like
Oppenheimer quite a bit to get used to Julian.
Pauli once referred to Oppenheimer's students as being Zunicker.
Somebody who knows enough German knows what this means---people who nod their
heads, and Julian wasn't that way---that and his hours.
However, he thought better of it and he soon learned to not only respect but
to love him and that is how it went, and then came the war.
Meanwhile, except for Fermi, and the explanation of the beta-ray spectrum, and
so on, nothing terribly much happened of a fundamental nature in physics
although there was the experimental discovery of the mesotron.

Well, the war came and Julian got sucked into it.
At first he started at Chicago in the metallurgical project and was persuaded
to come to Cambridge for the radiation lab.
I had nothing to do with it very much except that I approved of it and, well,
it seems that he came---he just came.
And weeks and months later, the garage kept on calling the university, what to
do with Julian's car?
They went to his room and they discovered numbers of uncashed checks.
He was a real scholar.
Now, from that rather rarefied region of meson physics and of course quantum
electrodynamics, came the very practical basic problems of the radiation
laboratory, the war laboratory in Cambridge.
And in this period, it was interesting to see, if you were leaving
approximately five o'clock in the afternoon, leaving your lab, there was a
single, solitary figure coming up against the current---in other words, you
could see that his life work was sort of perceptive in this way---against the
current.
He was coming to work.

A story, which although it sounds apocryphal is true.
They'd be working all day in this theoretical group headed by Uhlenbeck and
didn't finish by quitting time and there was this integral on the blackboard,
so when they came in the next morning everything was rubbed out except the
solution to that integral.
This was the gremlin working nights.
Julian lectured, twice a week, during that whole period on work in progress,
which generally was of the form of translation of wave guide theory into the
lumped constant theory and as soon as Julian made a breakthrough in one of
these things, these other bright boys around the place, like Dicke, and so on,
just invented all sorts of stuff.
It all went into weapons of some kind but they started with this breakthrough
of the translation of wave guide theory.
The radiation lab, at the end, published 27 volumes of the radiation lab
series.
Most of you who are old enough have seen those and studied from them.
There was all the lore of that period.
There was supposed to be a volume by Julian but the editor didn't know how to
handle Julian and didn't get this volume, so Julian's very fundamental work
there is missing.
I think some of it was published later by Marcuvitz.

And this, more or less, takes us up to that period.

Now a few more general remarks that will apply to many members of the
audience.
James Franck once told me---it was quite well known at that time---that there
are three stages in a man's life: werden, sein, bedeuten---the coming, the
being, and the signifying.
And well, theoretical physicists make a distinct translation---he is born, he
gets his doctorate degree, and then on an impetus works very hard, and by the
age of 30, he has the Nobel Prize.
Not in hand, not at all, but the work is there, he sort of knows it, but he
isn't sure.
Then there's a long stretch of time---to be or not to be, as Shakespeare
said.
Sometimes it takes---it took Einstein about 15 years, and Van Vleck, I'm not
good at mental arithmetic any more, but it took much longer than that, it must
be something like 40 years---but by the age of 30, give or take a few, for
theoretical physicists you've done it.
And the rest is consolidation, power, things of that sort.
And then comes a later period which gradually merges into the---I don't know
quite what to call it, a stuffed shirt would be too strong a term---but it's
something else, it's certainly not the same thing.

So this is an interesting celebration.
I was 60 once.
It's to some extent a difficult period in one's life.
There're not so many here who are there, but quite a few who will soon be
there.
So I will in general make a few remarks about suggestions if you're expecting
then to live a long life, and how to do it.
Number one, act your age.
You're crazy if you try to compete with the young fellows.
This is especially true for experimental people.
You don't have the quickness, the stamina.
You have something else.
But what is called productivity in the normal sense, you won't find.
You go through the careers of the great men of our era, Einstein, Bohr,
Heisenberg, Pauli, Dirac, Schr\"odinger, de Broglie, after the great surge,
you don't do it again.
Of course, you'd expect them to do more.
The daughter of one of my colleagues watched him blow a smoke ring, and she
said, ``This is an O, Daddy, now blow a T.''
He's not going to do it.

But it is terribly important to this mission which Julian has fulfilled, and I
don't know if he meant it on purpose, to sort of scatter his seed: 80 Ph.D's
under his direction.
I think that adds up easily to the combination of Bohr, Einstein, Dirac,
Pauli, Heisenberg, Schr\"odinger, de Broglie---and that's quite a challenge to
the future.
So I'm eagerly awaiting to hear the results.
Some of them are right here before me.
Some of the new ideas make me uncomfortable.
And that's as it should be.
It's in the direct tradition.
I know how uncomfortable Enrico Fermi was with the later developments of
quantum mechanics, and we all know how Einstein fought this gallant fight over
a great period of his life.
He wasn't at all lacking in ability, but what was happening in a sense, has
happened.
Now we're---I've jumped ahead of myself because before the philosophy part I
intended to put in, it's gone by---I've left out the greatest period of all.
The post-war period, the immediate post-war period.
When---sorry Van isn't here, I'd have to rub his nose in it, this sort of
thing---Columbia came to the rescue of New York, Julian and Dick---were the
experiments of Lamb and Retherford and Nelson and Nafe, which Julian
immediately interpreted as an extra moment of the electron and within a very
short time, just about the time when both of these gentlemen were 30, we had
the completion of the quantum electrodynamics as a finished business and at
the same time, the discovery of the meson---they came one after the other.

One of the new, challenging fields, and ushered in the title of the next talk,
``Schwinger and the Boom in Theoretical Physics,'' the big boom did come right
after that.
So I'm not going to make any peroration about Julian, except to say that to
me, my life with him was extremely \unclear\ and a great inspiration.
I've learned much of what I know from either fellow graduate students or my
own graduate students and one of my regrets, as I'm retired, is I don't get
any more graduate students to teach me, but it was a good time while it
lasted.
Again, final statement:  Don't overdo it.

\section*{Morton Hamermesh}
These are just to keep me company.
It's a great pleasure to be here and I thought on this occasion I might start
by telling you the title of this talk which the organization committee
refused, apparently, to print in the program.
I think it was called ``When We were One and Twenty.''
At least it was when I was one and twenty, he was younger.
I remember at City College, this was around 1934, I was a mathematics student.
I regarded physicists as an absurd bunch of people.
They always fiddled around making all sorts of strange approximations and they
were hardly what I call pure.
There was one person whom I knew very slightly and that was Julian who
apparently had been a student at Townsend Harris where he had as his teacher
Irving Lowen, who later came to NYU and, in fact, taught a number of us.
I know he taught me Dynamics and Relativity.
And then Julian arrived at City College and I knew him slightly until we
became students in a class together and I don't know if he remembers this, but
maybe he'll be sorry when I tell about it.
This was in 1934, '35.
We were in a class called Modern Geometry.
It was taught by an old dodderer named Frederick B., I think his name was
Reynolds.
That's it: Frederick B. Reynolds. Thank you.
Frederick B. Reynolds.
He was head of the Math Department.
He really knew absolutely nothing.
It was amazing.
But he taught this course on Modern Geometry.
It was a course in projective geometry from a miserable book by a man named
Graustein from Princeton, and Julian was in the class, but it was very strange
because he obviously never could get to class, at least not very often, and he
didn't own the book.
That was clear.
And every once in a while he'd grab me before class and ask me to show him my
copy of this book and he would skim through it fast and see what was going
on.
And this fellow Reynolds, although he was a dodderer, was a very mean
character.
He used to send people up to the board to do a problem, and he was always
sending Julian to the board to do problems because he knew he'd never seen the
course, and Julian would get up at the board.
And of course projective geometry is a very strange subject, the problems are
trivial if you think about them pictorially, but Julian never would do them
this way. 
He would insist on doing them algebraically and so he'd get up at the board at
the beginning of the hour and he'd work through the whole hour and he'd finish
the thing and by that time the course was over and anyway, Reynolds didn't
understand the proof, and that would end it for the day.
And that was my introduction to Julian.

The only other thing I remember from the time was that apparently Julian used
to help various of the young instructors at City College who were working on
their Ph.D. theses at Columbia.
The usual thing was, you taught at City College and you worked at Columbia and
he was sort of the unofficial adviser to various people and this well-known
habit of his of working late at night and not appearing until then, I'm sure
started then.
All sorts of people bugged him about problems they wanted solved and so this,
I think, was his means of keeping away from them. 

Then a little after that I remember that Julian disappeared and went to
Columbia.
Through the good offices of Rabi and some others, he went there, and he no
longer had to make classes.
Apparently that was the idea.
I went on and finished at City College and started at NYU and after about a
year I decided I was getting sick and tired of taking courses.
I wanted to do some research, except that my Professor, Otto Halpern, just
couldn't be bothered with his students until they'd really ripened and he
would leave us all alone: ``Go do something, don't bother me, I can't be
bothered.''

And at that time Hy Goldsmith, whom I had known at City College, and Julian
were at Columbia and I guess I talked to them and somehow or other---it was a
very mysterious process.
Columbia was an amazing place in those days.
Really.
A golden age for physics.
They didn't care whether you had a job there, whether you were a student
there.
You simply walked in and worked.
No one paid me, you understand.
I didn't need them to pay me.
I had an assistantship at NYU, but all you had to do was come in and you could
work.
I knew lots of people who worked at Columbia who had absolutely nothing to do
with the place.
You just knew somebody and so you came to the party.
And I remember very well that the first piece of research that I worked on was
something having to do with a paper that Herman mentioned today.
It was a paper on the widths of nuclear energy levels by Manley, Goldsmith,
and Schwinger, and I was looking for something to do and I came there and
Julian said, ``Well there is this thing that should be worked out and I
haven't done it because it takes a little more work. Here's what you do.''
It was some calculation about albedo of neutrons and I didn't know what he was
talking about.
I mean, I really was miserably educated.

And then a process began which recurred very often after that.
The calculation involved a great deal of information about Bessel functions
and in general, special functions, and Julian said, ``I'll show you how to do
the calculation,'' and he started off, ``now this is the differential equation
satisfied by Bessel functions and then here's how you get the solution.''
And then he went on this way for about, I think, it must have been about five
days.
He put in an enormous number of hours, by which time he had reproduced at
least three-quarters of Watson's \textit{Bessel Functions}, but this
apparently was a technique he liked to use.
It was very helpful to me because I began to see how you did physics.
At least mathematical physics.
And I remember the great pride I had when my name appeared in a footnote in a
paper in the \textit{Physical Review}.
I mean, it was my maiden effort.
And this was the beginning of a whole series of things, where we had a very
interesting method of working.

It was along about this time that I started to help Julian and Goldsmith and
Cohen, Bill Cohen, in some further experiments in nuclear physics, including,
of course, the neutron-proton interaction that Herman mentioned and---life was
very strange.
I would work up at NYU or City College, come to Columbia around three o'clock,
start doing calculations.
Julian would appear some time between four and six and we would have a meal,
which was my dinner and his breakfast, and then we would begin the evening's
work, which was a strange combination of theoretical and experimental work.
We were experimenters, if you can call us that.
That is, we were capable of putting foils in front of a radon beryllium source
and measuring transmissions through them and activations, like grabbing the
foils, running down the hall of Pupin---it was on the top floor---running like
crazy, putting the foil on a counter, and taking a reading.
And then we would run back, put them up again, and start doing theoretical
work.
And we would work rather strange hours.
It seemed to me that we would work usually to something like midnight or
\mbox{1 a.m.}, and then go out and have a bite to eat.
This would mean two or three hours during which I would get educated on some
new subject.

\newpage

I learned group theory from Julian, and I must admit I forgot it all
immediately, but as I recall, I had all of Wigner's book given to me, plus a
lot more at the time and this was a regular process we went through and I
think this must have gone on for a year or so, and we started doing
calculations on ortho-paradeuterium and on ortho-parahydrogen, scattering of
neutrons, and this involved just an unbelievable amount of computation.
And we would work on this nights and there was this wonderful theoretical
seminar at that time in New York.
There was a joint theoretical seminar of Columbia, NYU, and anybody else in
the City who could come to these things, and it was a sort of a battlefield.
My professor, my official professor, Halpern, would take on anybody and he was
a rather testy fellow and loved getting into arguments. 
He would just take the greatest pleasure in taking on Gene Feenberg, or Fermi,
when Fermi came, or anybody else.
It just didn't matter.
There were just violent fights.
I can recall giving a seminar there where, it seems to me, I prepared for this
seminar for a month---a little bit with Julian's help---and then started to
talk and I said something and Fermi objected and Halpern came to my defense,
and as I recall, I never got another word in edgewise.
Never spoke another word through that hour.

Well, the work went on for a while and we got all these computations done,
except that this was a period, as I recall it was around 1938, beginning of
'39, and I think Julian was getting ready to go off to Berkeley, and the paper
was done and we were going to write it up and I looked upon this as my magnum
opus.
You know, I was going to be doing a thesis with Halpern, but who cared about
that.
This was really great stuff.
Then we started to write the paper.
The only trouble is that at this time Julian was already very much interested
in the tensor forces, and I remember very well helping him with some
calculation involving the coupled differential equations that you get.
And I was a great reader of the literature and I was always telling about
interesting problems and unfortunately one day I mentioned the absorption of
sound in gases and that started him off on an enormous amount of work which I
don't think he ever published, as far as I can tell.
But he did all sorts of calculations on this and there I was, trying to get
him to write a paper and he's a rather finicky writer---maybe he isn't so
finicky any more---but I can recall that there were only a few weeks left
before he was to leave and there was the paper and we were still in the first
paragraph and every night we would start, we would write six or seven lines,
and we wouldn't get it done, and here I could see the time slipping, and I
would go home and I would cuss hell out of him---to myself.
And at one point I contemplated murdering him, but I didn't.
He went off to Berkeley, paper not done, and then in 1940, suddenly there
comes a telegram, ``Please send all the calculations,'' and I packed up a pile
of stuff about this high, shipped it off, heard nothing till suddenly some
letters appeared in the \textit{Physical Review}.
There were some experiments by Alvarez and Pitzer, and a short note by Julian
with the calculations, and I just gave up on him, did a thesis quick, got it
done and then, in 1941, I suddenly discovered I had a job through the
character.
He actually had managed to find a job for me at Stanford.
It seems to me he was at Berkeley and I believe it was he who talked to Felix
Bloch.
No? OK, then I'm under delusions.
I've given him credit for my job.

But the next time I saw Julian was in Cambridge.
I came to the Harvard Radio Research Lab in '43 and Julian arrived there about
the same time, at the radiation lab, and we saw each other and he said to me,
well, you know, we really ought to write that paper.
That's a great idea.
It turned out, of course, he really had a point.
He had found a very neat trick for reducing all this unbelievable amount of
calculation that we had to do to what then amounted to four days of work and
so we did it all over again very, very quickly and the paper was finished in
about two weeks, I think, of writing.
He had improved his style by then and it was published, I think, in '46 and
then another one in '47.
Well, essentially what I'm trying to say is that I think I should claim that
I'm Julian's first student.
I believe I learned more from him than I learned from anybody else.
In fact, I think he's the only one from whom I ever learned anything.
I find it very hard to learn from other people, but in his case I would say he
showed he was a great one-on-one teacher as well as fine lecturer.
It's really a great pleasure to be here.
There are lots of other people in addition to Julian, whom it really gives you
great pleasure to see.
There's a certain special pleasure in seeing people whom you've known for a
long time, with whom you've worked, with whom you've had miserable troubles
and, on the other hand, gotten great pleasure.
For one thing, it reminds you that you were once young and I  hope there  will
be lots more reunions of the same kind.
Thank you.

\section*{Bernard Taub Feld}
Well, I must say that I feel like the aging vaudeville comedian who comes on
after a couple of acts that are impossible to follow.
What with Rabi's reminiscences about the good old days of physics in New York
and then Morty's reminiscences about doing physics in the mid to late 30's at
Columbia with Julian around, all I can add to this picture is to note that I
was around more or less at that time and I was going to talk---and I guess I
still have to talk---briefly this evening about some aspects of life growing 
up as a physicist or trying to become a physicist in New York in that period.
But I must say that my view is a somewhat different one from, certainly from
Rabi's and to a large extent from Morty's.
You might call it a worm's eye view of growing up in New York in that period.

I was an undergraduate at CCNY a few years after Julian had already become a
legend and by the time I got to Columbia, Julian was just taking off for
Berkeley, so I followed enough afterwards to have absorbed a great deal of the
legend which Julian very rapidly became in New York, although fortunately, I
did get to know Julian somewhat in that period and I guess I will talk a
little bit about that.
You've heard a number of stories about actually people who worked in New York
in that period but I guess there was something very special about the period
of, let's say, the 1930's.
Not only in New York but in the United States, as far as physics was
concerned.
It was really a period when the United States, when physics
in the United States developed at a fantastically rapid rate.
I mean, before, let's say in the early '20's, there were a few isolated
instances of good physics being done in the United States.
There was of course Michelson, and Millikan, and Rowland, but you could count
them on the fingers of one or two hands.
They were great experimentalists, but it was really not in the mainstream of
modern physics.

In a certain sense, the United States was dragged into the mainstream of
modern physics in the late '20's as a consequence of the efforts of a few
people.
These were young physicists who obtained their Ph.D.'s in the United States
and decided that what was really going on of interest was all going on in
Europe---the development of the new quantum mechanics.
And they somehow or other got themselves fellowships and went off to Munich or
to Zurich or to Gottingen and they worked with the greats who were then
developing---or to Copenhagen---who were then developing quantum mechanics and
came back determined to introduce the modern physics into the United States.
We're fortunate to have one of them with us tonight.
Rabi was certainly one of these pioneers.
In fact, Hamburg, too. Yes. I had it written down here but I wasn't reading.
I would say that modern physics in the United States owes a tremendous debt to
two people.
Two of these pioneers---Rabi was one, Robert Oppenheimer was the other---who
came back determined that they were going to start schools of modern physics
in the United States.
Oppie went off to the West Coast and started a school of theoretical physics
at Berkeley and Pasadena, and Rabi settled in New York and turned New York in
the period of the '30's, or was certainly one of the people who was primarily
instrumental in turning New York in that period, into what one might refer to
as a kind of experimental Copenhagen.
I don't know whether that's the best description but it certainly was a center
where young people could come and could do physics.

Now there were a number of things that happened in New York in the '30's
which, of course, contributed---at least three.
Two of those were in a sense historical vicissitudes and in fact, really, they
represented very tragic events in world history and by a kind of what you
might call a principle of the ill wind, the effect in New York, or in the
Eastern part of the United States and to some extent all over the United
States, was to turn, was to build up modern physics in the United States. 
One of those was the Depression, the Great Depression of the '30's and the
other was the horror of the development of Nazism in Germany.
Consequence of the Depression was that there were in New York at that time, in
that period, a fairly large number of young, aspiring, bright, extremely
bright physicists, or aspiring physicists, who needed jobs and were willing to
work long hours and low pay and there were lots of jobs available in the---or
at least a fair number of jobs available at that time---teaching physics in
the City College system in New York.
In CCNY, in Brooklyn College and Queens College and Hunter College, and they
were able to get jobs.
If I recall correctly a normal teaching job in New York at that time was
something like teaching 20 or 25 hours a week and the pay was something like 
\$1500. 
At that time that was regarded as pretty good---\$1600, was it?---and a number
of people in this room did that.
Now, the other thing about New York was that there was Columbia University and
Rabi was there and the head of the department, George Pegram, was a very
farsighted man and with, I suspect, a certain amount of prodding from Rabi,
Columbia took advantage of the presence of all those people and as Morty just
pointed out, any aspiring, bright, young physicist could do research at
Columbia.
There was Rabi's lab, where there was a cyclotron, and all these people
gathered around Columbia to do research and gathered around them all the
bright people they could find.
So this was a special period and the final aspect was that there was in New
York City at that time this reservoir of young, aspiring, mainly children of
immigrant parents, upward mobile, with a kind of tradition of learning and
respect for learning and hard work, who were very anxious to take advantage of
this situation and who were the fodder out of which the physicists were
turned, so it was this combination of things.

Now, to be absolutely fair, there was more to physics in the United States
than just New York and vicinity and the West Coast.
There was a great hinterland, which most New Yorkers thought of as a sort of
prairie, that started out on the other side of the Palisades and then ended up
in the Rockies, which was mainly useful for producing corn and steak and
hay fever in the summer, but there were some things going on there.
In Chicago, for example, there was the Compton effect and there was the school
of work in cosmic ray physics, which was quite important, and then there was
Michigan which, at least in New York at that time was regarded as a sort of a
glorified summer colony in physics where the great Europeans would come in the
summer to lecture and if you were lucky you could get to go and listen to
them, and they had Goudsmit and Uhlenbeck, and Fermi came to lecture, and so
on. 
That was a great place. 

But after that there was Berkeley and Pasadena and New
York and surroundings---and surroundings meant, oh, Princeton and then Cornell
and the other aspect, of course, was that all the great European physicists
who fled from Nazi Germany got funneled through New York and some of them
stayed and some of them stayed long enough to give some courses in New York.
Teller and Bethe, who didn't stay in New York, nevertheless, as I recall, came
back in this period and taught quantum mechanics to graduate students at
Columbia and in any event---of course Fermi and Szilard, when they came they
just stayed, at Columbia, but everybody passed through so that for a student
at that time it was really a great paradise.
Well, that's more or less the atmosphere into which Julian flashed in the
early '30's and where I came along a few years later.
And as I say, by that time there were a certain number of legends surrounding
Julian.
It's amusing to recall some of these legends because they are legends which
refer to some of the same things that Rabi was telling us about and you got a
first-hand account this afternoon of some of these stories, but when I tried
to recall them in the last few weeks, trying to think of what was going on in
New York and my recollections of the stories about Julian, the stories were,
somehow or other, just slightly different.
It's interesting to compare the legends which didn't take long to grow up---as
I say I came along just a year or two after Julian. 

For instance, there was a legend at CCNY that Julian was first discovered by
Irving Lowen.
Irving Lowen was then teaching at Townsend Harris, and the story is that
Irving came across this kid sitting in the library reading the
\textit{Physical Review\/} and he looked over his shoulder and there was this
kid reading Dirac and so Irv thought, well, here's another one of these
smart-aleck kids that, you know, we get them every once in a while, so he
quizzed him about what he was reading and Julian allegedly was not only
capable of telling him what he was reading but also told him what needed to be
done to complete what Dirac hadn't completed in this particular paper. 

The other legend---or rather the story you got from Rabi---was how Julian
shifted from CCNY to Columbia.
Now the story that went the rounds at CCNY at the time when I just got there
was a little different.
It had to do with a chairman of the department, a man called Corcoran, who
was---well, he was not the greatest physicist who had ever been around.
A rather crotchety old Irishman, and allegedly when Lloyd Motz had taken
Julian under his wing, he took Julian to Corcoran and said look here's this
kid and he really knows more physics, too much physics, to have to take these
elementary physics courses.
Why can't we let him take some of the advanced courses and Corcoran is alleged
to have said, ``Over my dead body.
As long as I'm chairman of this department, no smart-ass kid is going to be
allowed to skip taking my course in elementary particle physics.''
In elementary physics, excuse me.
So Motz was supposed to have taken Julian to Rabi and Rabi straightened it all
out by getting a scholarship to Columbia.
That was the story.
Now we heard the real story from Rabi but it's interesting to compare some of
these stories. 

There was another story which parallels the one Rabi told, a story about
Julian's taking a course with Uhlenbeck. 
That wasn't the way the story went when I got into Columbia.
The story had to do with a chemistry professor and I'm not sure I remember his
name---Kimball, I think, LaMer.
This was a chemistry course that Julian was supposed to have had to take.
And now, the story was, it is probably completely apocryphal---that Rabi got a
telephone call one day from Professor LaMer who said, ``This kid, this
prot\'eg\'e of yours, this Schwinger, I'm going to have to fail him.''
And Rabi said, ``Why?''
``Well, he's taking my course and he hasn't appeared at a single lecture all
term.''
And Rabi was supposed to say, ``Well, why don't you fail him?''
To which LaMer was supposed to have answered, ``The trouble is, he just took
the final and he got 100 on it.''
And now, Rabi is supposed to have answered, ``Well, look, are you a man or a
mouse? If you want to fail him, fail him.''
Besides, the story is supposed to go on, Rabi said, ``How can you pass up this
opportunity to go down in history as the man who failed Julian Schwinger!'' 

Well, now, these were the apocryphal stories that were going the rounds when I
arrived on the scene.
I got to CCNY in 1935 and actually when I came to CCNY as a freshman, I
thought I was going to be majoring in history.
This came about for a number of reasons, one of which was that I had a great
memory for dates, and history in those days, at least in the high school
courses that I took, was the memorization of a long series, a semi-infinite
series of dates and since I was great at memorizing it, I thought I was
obviously a born historian.
So I thought I was going to be a history major but I also had an idea that
maybe science would be fun and so I had taken a course, I'd taken
physics---this was at Boys High School, which was a rather good high school,
but the man who taught physics was not really much of a physics teacher.
I remember two things.
One, his great pedagogical tour de force was he would sometimes if it was a
nice, cold, crisp winter day, he would go shuffling across the room and then
open a bunsen burner and light it by a static electric discharge from the tip
of his finger and he thought that this was really, you know, the way you
taught electricity to a bunch of high school students.
This was great physics.
The other thing is I got an ``A'' in the course, but the reason I got an ``A''
in the course was that the final exam contained a lot of simple questions but
the one question that everybody else failed on, which I got like that, was the
question of ``Name three forms of Ohm's Law.''
I had no difficulty in naming the three forms of Ohm's Law.
Some of you may not know what the three forms of Ohm's Law are, but I still
remember them very well.
They are: $V=RI$, $I=V/R$, $R=V/I$.
That was passed for physics, so physics to me was not a very interesting
subject.
Then I came to CCNY and then for some reason or other---I'm not sure whether
it was required or whether I just thought it might be fun to try it again
anyway.
Anyhow, I took elementary physics and then I had real luck because my first
instructor in elementary physics was Morty Hamermesh.
He was just a graduate student then and he was teaching elementary physics at
CCNY and all I can say is that after that term I was no longer a history major
and after that sort of I felt that my \unclear.

And the second thing was that I met Hy Goldsmith and this was right in the
middle of the period that Morty was talking about.
In my sophomore year I was taking a course with Hy, and Hy was a dilettante and
he was a very interesting guy because he was very lazy.
But on the other hand he had two things---one he had an encyclopedic knowledge
of what was going on and he had really good taste. 
He knew what was important and what wasn't important.
But at that period, I guess, it was just a period I think, probably when Morty
and the other people working with him at Columbia had gotten to be pretty
tired, gotten pretty tired of running up and down the hall with the foils and
so I was recruited, as a sophomore then, to do the running.
I guess Marty doesn't remember but I spent six months at Columbia doing the
sprinting.
I was a pretty good sprinter.
I didn't know anything else but they were studying resonances in rhodium, and
rhodium---I've forgotten what the mean life is now, but it's really very
short.
You had to take those foils and sprint that 50 yards from the irradiation to
the Geiger counter and I was the fastest sprinter they could find.
I was a real good sprinter then, so I made out real well.
As a result of that I not only got to hang around at Columbia at night but
even when they went up to see Julian to consult on the theory or when
something had gone wrong with the experiment and they got bored and just went
up to talk to Julian, I was allowed to go up with them, and so I got to
listen.  

\smallskip

(From the audience: ``Did they pay you by the trip or the minute?'')

\smallskip

Listen, the payment was---I was allowed to go along and that was great stuff,
you know.
A sophomore in college doing that.
That was really being a student.
Well, that was my introduction and as I say, I sort of got to know Julian
because I was allowed to sit in on these sessions when they worked out the
theory and in that period, I think, Julian was really on the way to becoming
an experimentalist.
You never saw him running down that hall, but I just looked up the Physical
Review the other day and in the period from '37 to '39, Julian published six
papers and three of those were experimental and I suspect---I don't know, I've
never asked Julian---I have a suspicion that one of the reasons that he ran
off to Berkeley---there probably were other reasons in '39---but one of them
was to get away from these mad experimentalists who didn't give him any peace,
who would come wandering up at all hours of the night to disturb his work and
who were trying to turn him into an experimentalist---God forbid.
But in any case, by the time I got to Columbia, Julian had left, had just left
and I didn't meet Julian again until 1943, when I was in Chicago and Julian
arrived at the metallurgical laboratory to spend something like six months, I
guess it was---two months. 
Anyhow, it was in the summer, a few months in the summer of '43.
I'd been there a year and I was pretty sophisticated by then.
In fact, I was doing experiments.

That was the period in Chicago after the chain reaction had been proved, and
what the people in Chicago were doing was designing the Hanford Reactor.
Now that was not easy because nuclear engineering didn't exist.
They were really discovering and inventing nuclear engineering.
This was Wigner and Szilard and Fermi and Wheeler and a crowd of young people
who were working with them.
They were all physicists but in a pinch a physicist, a theoretical physicist
make pretty good engineers.
In fact, they were all first-class engineers.
No question \unclear\ and Julian turned out---I didn't realize that he'd been
doing engineering already at the radiation lab, but, Julian, as you heard this
afternoon, is also a pretty good engineer and he demonstrated it that summer.

I guess there were a number of problems that remained to be solved, mainly to
help in the design of Hanford, and these were in transport theory and some of
them were more or less difficult and Julian was working on some of them and
the things I was doing were not really directly connected with the Hanford
thing and the Hanford thing was the most, the design of Hanford was the most
important.
Anyhow, there was the need of some kind of a match between Julian and the
metallurgical laboratory, because Julian didn't come in until some time in the
evening and most people were already gone, so the contact wasn't well
established, so I sort of elected myself in that period to be the matchmaker.
I had known Julian from Columbia, slightly, but at least well enough that I
figured that I might be able to help in making the match between Julian and
the project.
It was really a very interesting process because what would happen was that I
would go around in the afternoon---not every afternoon, but occasionally---go
around to my friends in Wigner's group and sort of try to smell out what were
the problems with which they were having trouble.
Things that were giving them some difficulty.
And then some time in the late evening, maybe 10 o'clock or 11 o'clock or
something, I would wander off to Julian's office and wander in and Julian
would be sitting there at his desk, typically, this was a very hot summer in
Chicago, and Julian was a very fastidious dresser in those days.
He never took off his coat.
In Chicago I never saw Julian without his jacket.
He'd be sitting there with his white shirt and tie, tie never even loosened,
jacket on, with a pad and a paper, he'd be scribbling furiously, working on
some problem on the pad, with his handkerchief, supersaturated handkerchief in
the left hand, mopping the sweat off his brow as he worked, and I used to
wander in and sit down, and wait and at some point Julian would pause to catch
his breath and I would kind of interrupt him and try to get his attention away
from whatever he was doing and I usually succeeded and not only because I was
a pretty persistent guy, but because Julian is a nice guy and if you sort of
bother him, he'll pay some attention to you, and after a while I would get him
interested in the particular problem I had in mind.

I'd start talking about it and Julian would get interested and then he would
go to work on it.
He'd get up to the blackboard and I would start taking notes.
As Julian worked on the problem, I would be taking notes and sometimes, you
know, that could be pretty hectic.
I don't know any of you who saw Julian work in those days.
Julian is ambidextrous and he can, he has a blackboard technique that uses two
hands, and frequently, when he really got carried away, he would be solving
two equations, one with each hand, and trying to take notes could be a hectic
job.
Well, at some point, either we would finish the problem or the dawn would
start to break in the eastern horizon, and we would decide it was time to quit
and then often, we would go to have breakfast together.
We would get into Julian's sleek black Cadillac and go to the nearest
all-night eatery, where we would both have breakfast of, I think, it was a
steak.
So we would have our steak breakfast and then off we would go to our
respective beds.

Now this wasn't all that either I or Julian did that summer, but I suspect
that many of the knotty problems which the Hanford engineers weren't able to
cope with, were solved in that summer as a result of that particular
technique.
Well, after a while Julian went off to the radiation lab, to get back to the
radiation lab, and I went off to Los Alamos and we didn't meet again till we
both arrived in Cambridge after the war, but that history is one that
everybody is familiar with, so I guess that this is a good point to stop my
worm's eye view of the good old days when physics was really bursting out in
New York.
I tried to think of a title for this talk and the only title I could think of
that really fits it, is ``Some Reminiscences from a Well Spent Youth,'' I
think.

\section*{Richard Phillips Feynman}
(D. Saxon: And now it seems to me very important at this stage, in the
interest of accuracy and versimilitude that we have a New York physicist who
actually speaks like one: Dick.)

Not only that, but I thought this was Schwinger's birthday party, not Columbia
University's.
In 1935 I wanted to go to Columbia University and I applied there.
It was during the Depression and it was hard to get money, and they said yes,
but you have to take our examination and it will cost you \$15 and if you
don't pass, we keep the \$15.
I did not pass, they kept the \$15, and I've been cheated by Columbia ever
since.
Instead of getting a good education, I had to go to MIT while the expert was
there.
Therefore, it was not until I went to Los Alamos that I got a chance to meet
Schwinger.
He had already a great reputation because he had done so much work as the
others have described and I was very anxious to see what this man was like.
I'd always thought he was older than I was because he had done so much more,
at the time I hadn't done anything.
And he came and he gave us lectures.
I believe they were on nuclear physics.
I'm not sure exactly the subject, but it was a scene that you probably have
all seen once: the beauty of one of his lectures.
He comes in, with his head a little bit to one side.
He comes in like a bull into a ring and puts his notebook down and then
begins---and the beautiful, organized way of putting one idea after the other,
everything very clear from the beginning to the end.
You can imagine for a lecturer like me, what a sensation it was to see such a
thing.
I was supposed to be a good lecturer according to some people, but this was
really a masterpiece.
Each one of the lectures was a great discourse while what I did was a talk on
something.

So I was very impressed and the times I got then to talk to him, I
learned more, and then we went off and met many times in the different
conferences, which always was a pleasure, but the greatest conference was the
one at Pocono, I think it was, where we tried to describe---each of us had
worked out quantum electrodynamics and we were going to describe it to the 
tigers.
He described his in the morning, first, and then he gave one of these
lectures, which are intimidating.
They're so perfect that you don't want to ask any questions because it might
interrupt the train. 
But the people in the audience like Bohr, and Dirac, Teller, and so forth,
were not to be intimidated, so after a bit there were some questions.
A slight disorganization, a mumbling, confusion---it was difficult.
We didn't understand everything, you know.
But after a while he got a good thing.
He would say, ``Perhaps it will become clearer if I proceed.''
So he continued this, continued it.

\enlargethispage{1.0\baselineskip}

Then I was supposed to go on in the afternoon, and Bethe, Hans Bethe said to
me, ``Look Feynman,'' he says, ``notice every time Schwinger tries to explain
a physical idea how it gets all wrapped up in it, so you can't explain
anything to these guys, you better make it mathematical.''
Well, Schwinger could make it mathematical.
I only thought I could make it mathematical, so when my time came I started
out by trying to make an equation that would do everything.
They wanted to know where that equation came from.
So then I'd start to describe physical ideas and things went very badly.
Dirac would say things like---first of all I must explain our methods were
entirely different, as you all know.
I didn't understand about those creation and annihilation operators.
I didn't know how they worked, that he was using, and I had some magic way,
from his point of view, and Dirac would always interrupt me saying, when I was
explaining about how I was going to work out positrons and so on, ``Is it
unitary?''
And again, I'd say, again, well, I remembered his trick and I said, ``Perhaps
it will become clear as we proceed.''

But Dirac was not put off, and like the Raven kept saying, ``Is it unitary?''
And not being quite sure what unitary meant, I said, ``Is what unitary?''
So he said, ``The matrix that carries you from the past to the future,'' and
then I was saved because I said I had no matrix to do that because some of my
positrons are already in the future.
Like coming in from the wrong end.
And then I tried to go on and I explained how you could disregard the
exclusion principle in intermediate states and then Professor Teller said, ``You
mean that Helium can have three electrons in the S state for a little while?''
I said, ``Yes.''
That was chaos, and then, all the time I was pushed back, away from the
mathematics into my so-called physical ideas until I was driven to the point
of describing quantum mechanics as an amplitude for every path, for every
trajectory that a particle can take there's an amplitude, and Professor Bohr
got up and explained to me that already in 1920 they realized that the concept
of a path in quantum mechanics---that you could specify the position as a
function of time---that was not a legitimate idea and I gave up at that
point.

\enlargethispage{1.0\baselineskip}

But the thing that I really, why I told this story, is that right after that,
we got together in the hallway and although we'd come from the ends of the
earth with different ideas, we'd climbed the same mountain from different
sides and we could check each other's equations.
We compared our results because we had worked out problems and we looked at
the answers and kind of half described how the terms came.
He would say, ``Well I got a creation and then another annihilation of the same
photon and then the potential goes~\ldots''
``Oh, I think that might be what,'' I'd draw a picture that looks like this.
He didn't understand my pictures and I didn't understand his operators, but
the terms corresponded and by looking at the equations we could tell, and so I
knew, in spite of being refused admission by the rest, by conversations with
Schwinger, that we had both come to the same mountain and that it was a real
thing and everything was all right, so then after that we had many
conversations together, at Rochester Conference, and so on, and he's become,
and I've always had a great pleasure talking to him and with his gentleness
and sense of humor, and so on and so on, I'm very happy to be here tonight to
wish him a happy birthday and also to wish that we can continue our
collaboration as we have been doing before because I'm kind of working on
quantum chromodynamics.
My work to produce high energy results and I'm getting a bit confused and I
need the kind of elegance and style that Professor Weisskopf mentioned, and
I'd like to talk to you about it some time, so after this stuff's over, I'm
going to come back again and tell you.
Thank you very much.
Congratulations, Julian.

\section*{Closing remarks by David Saxon and  Isidor Isaac Rabi}

\noindent\underline{D. Saxon}\newline
Now I would like to finish with a few rather brief remarks.
Perhaps I'll begin by adding to Professor Rabi's rules for a long and
satisfying life an additional rule which I think is equally important to the
one you enunciated, Professor Rabi \ldots 

\bigskip

\noindent\underline{I.~I.~Rabi}\newline
Before you finish I'd like equal time on that Rabi story.
Julian was an undergraduate at Columbia, as I explained, and he suddenly
leaped from a student of low standing, precarious standing, at City College,
to getting a Phi Beta Kappa at Columbia.
It's not that our standards were lower than the City College's, although I
assure you, their standards were very high.
But something happened in this transplant.
Anyway, Columbia at that time had anticipated the matrix mechanics and
students were classified in a matrix with two indices.
One was ordinary credit and the other was maturity credit.
And you had a certain weight of ordinary credits and a certain weight of
maturity credits.
One Sunday morning I was called up by the dean, Dean Hawkes, and he said, ``What
shall I do about Schwinger?''
I said, ``What's the problem?''
He said, ``He has enough credits to graduate but he hasn't enough maturity
credits.''
It seemed to me absurd. How can you talk about things that way?
So I said, ``Well, you have your rules. I don't know what you can do about it.''
I wasn't going to make a great plea. See how the thing'd work.
Well, he was a real man, and on Sunday, he was a religious person, he said,
``I'll be damned if I won't let Schwinger graduate because he doesn't have
enough maturity credits.''
Of course this gave me great faith.

And then LaMer came along.
He was, for a chemist, awfully good.
A great part of his life work was testing the Debye--H\"uckel theory rather
brilliantly.
But he was this rigid, reactionary type, as described.
He came along, he had a mean way about him.
He said, ``You have this Schwinger? He just didn't pass my final exam.''
I said, ``He didn't? I'll look into it.''
So I spoke to a number of people who'd taken the same course.
And they had been greatly assisted in that subject by Julian.
So I said, ``I'll fix that guy. We'll see what character he has.''
``Now, Vicky, what sort of guy are you anyway, what are your principles?
What're you going to do about this?''
Well, he did as said flunk Julian, and I think it's quite a badge of
distinction for him, and I for one am not sorry at this point, they have this
black mark on Julian's rather elevated record.
But he did get Phi Beta Kappa as an undergraduate, something I never managed
to do. 

\bigskip

\noindent\underline{D.~Saxon}\newline
Thank you, Professor Rabi.
Earlier today you were commenting on some rules for a long and satisfying
life, and I want to add one to it which I expect most people here will agree
about, and that is you only ought to talk about things you really know
something about.
Not all of us observe it, of course.
I intend to take my own advice very seriously and therefore I'm not going to
talk about physics but instead about universities, which is my present field
of specialization. 

\smallskip

(From the audience: That means you know anything about it?)

\smallskip

The hard way. The hard way.
One of the most difficult problems that we who have responsibility for
university administration have to contend with, is an astonishingly forceful
tide of anti-intellectualism, which takes many forms.
I'm not going to try to run through all of the manifestations of that
phenomenon, but there's a particularly troublesome one which is the repeated
contention that in our great universities, research somehow comes at the
expense of teaching.
That scholarship only happens if people neglect teaching.
Now I believe that's manifestly false, but it's been extraordinarily difficult
to persuade people of that fact and therefore I was especially struck by the
fact that earlier today both Professor Weisskopf and Professor Rabi called
attention to that dimension of Julian's contribution.
Professor Weisskopf mentioned that there is something like 100 Ph.D. students
of Julian's, Professor Rabi, 80.
The most striking thing about that is the difference between 100 and 80 is 20,
which is more than most people have under any circumstances.
Now, it so happens that probably never before in my life have I had a chance
to get one up on Professor Weisskopf.
Never before in my life have I had a chance to get one up on Professor Rabi.
And to do it to both at the same time is really remarkably improbable, but I
performed an experiment.
I'll tell you what I did.
I asked Julian and I know the answer.
The answer is 67.
But that happens to be an aside.
I couldn't resist it.

The first remark to be made is that Julian, in addition to all the work he did
in physics, and incidentally, the unpublished work equals in volume, or
perhaps exceeds in volume, the published work.
I know that for I've seen a very substantial part of the unpublished work.
Let me also mention that if you want to talk about Julian's students, you need
also to talk about his collaborators because his collaborators have almost
always also been his students.
Morty referred to that.
Morty was a collaborator of Julian's, not a formal student of his.
I was a collaborator of Julian's, not his student in the formal sense.
The number of people I know---Rarita, Gerjuoy, Marcuvitz, Frank Carlson,
Harold Levine, John Miles, Albert Hines, and I suspect, to some degree, Edward
Teller with a 19 year old Julian and Isador Rabi with a 17 year old Julian,
collaborators of him were also in some way his students.
But in addition to that, as has been pointed out, the auditors in his courses,
from other institutions often---Black, Jackson, and so on---were also his
students.
I think the message here is a very clear one to me: that great physicists,
great scholars, are also great teachers.
I expect almost all of you in this room would agree that that was the case.

Why is it that we are so little believed when we try to make that point?
Why?
Why does that statement fall on deaf ears when all of us live it and know it.
Perhaps, may I suggest, because we have not been willing enough to inform
others about our subject.
Not merely for their good, but also for our own.
I'm sure each of you will have some thoughts on that problem but I expect each
of you is as far from a solution to it as I am.
I am desperately in need of a solution.
That's a problem that's ever present with me and it is worth serious thought,
let me assure you. 

Well, let me conclude by remarking that we're celebrating a great occasion
this weekend.
We're celebrating it at UCLA which gives me very special pleasure.
Vicky, I have to say that because even though you're unhappy that Julian left
Cambridge, I am convinced beyond any argument that I am largely responsible
for the fact that he's at UCLA.
I believe it to be true and therefore it gives me very special pleasure.
Celebrations such as this, I think, are occasions which convey very much mixed
emotions.
On the one hand, the joy of recognizing a life of extraordinary
accomplishment, the joy of being again with old friends, but also there's the
bittersweet taste of inexorable passing of the years, not only for Julian but
for us all.
Thank you very much.


\newcommand{\unknown}{\textsl{Nomen Nescimus}}

\begin{landscape}
\centerline{\bfseries %
  Participants in the SchwingerFest celebrating %
  Julian Schwinger's 60th Birthday (UCLA, February 1978)}
\par\bigskip\par
\centerline{\includegraphics[scale=0.672]{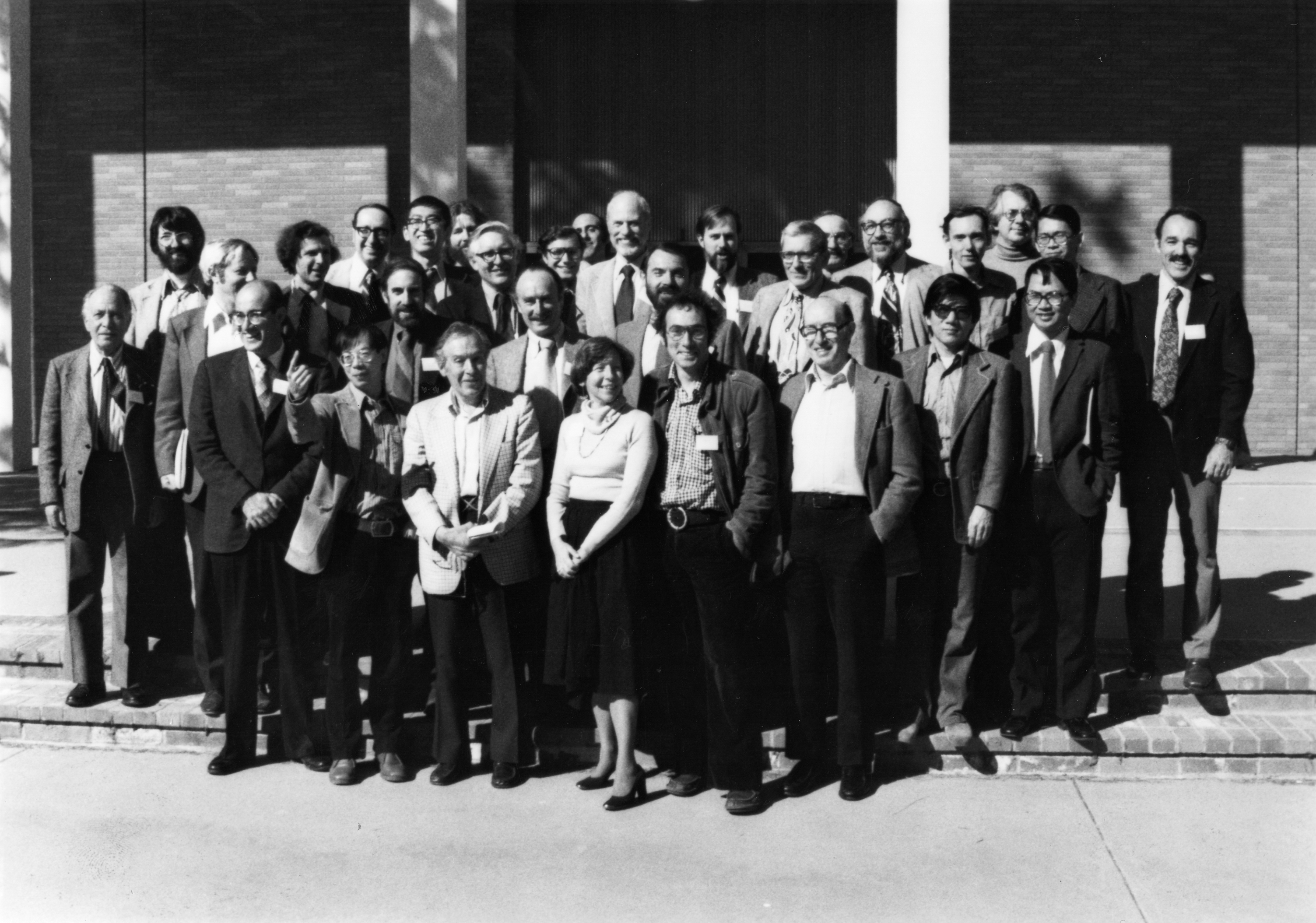}}
\end{landscape}

\begin{landscape}\vspace*{8.5mm}\noindent\hspace*{-0mm}%
  \begin{tabular}{@{}c@{}}
\includegraphics[width=450pt,viewport=100 120 730 440,clip=]%
    {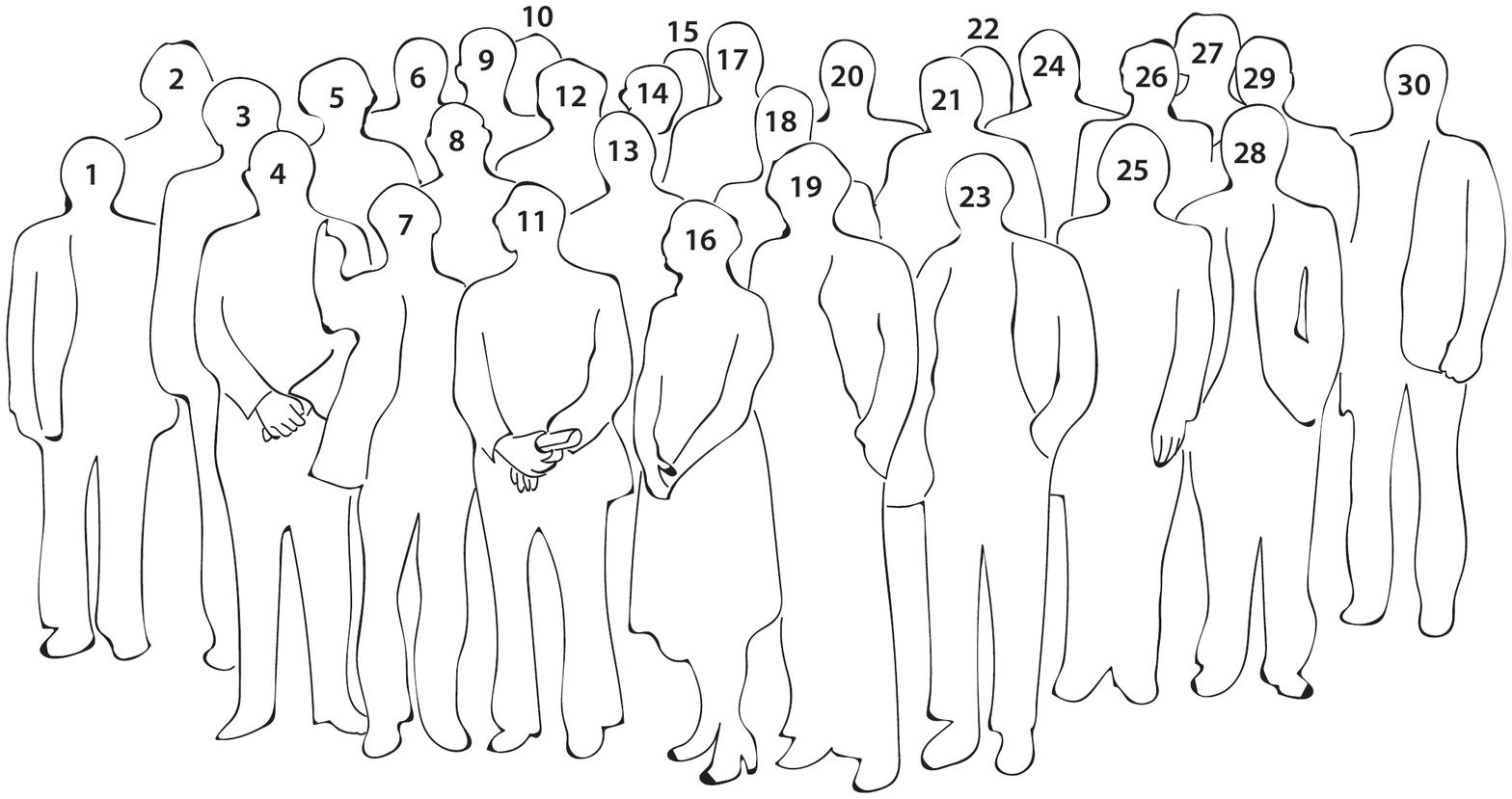}\\ \small
 \begin{tabular}{@{}r@{\enskip}lr@{\enskip}lr@{\enskip}l%
                  r@{\enskip}lr@{\enskip}l@{}}
  [1]& William Rarita & [7]& Y. Jack Ng  &[13]& Roger G. Newton 
                    &[19]& Gordon Baym &[25]& York-Peng Yao \\{}
  [2]& Kimball A. Milton & [8]& Charles Sommerfield  &[14]& Alain Phares 
                    &[20]& David G. Boulware &[26]& Lester L. DeRaad, Jr. \\{}
  [3]& Paul Martin & [9]& Shau-Jin Chang &[15]& \unknown
                    &[21]& David Saxon &[27]& Sheldon L. Glashow \\{}
  [4]& Charles Zemach &[10]& Donald Clark  &[16]& Margaret Kivelson
                    &[22]& \unknown &[28]& Hwa-Tung Hieh \\{}
  [5]& \unknown &[11]& \textbf{Julian Schwinger}  &[17]& Bryce DeWitt 
                    &[23]& Bernard A. Lippmann &[29]& Lai-Him Chan \\{}
  [6]& Michael Lieber  &[12]& Kenneth Johnson &[18]& Stanley Deser 
                    &[24]& \unknown &[30]& David D. Lynch 
  \end{tabular} 
\end{tabular}
\end{landscape}

\end{document}